\documentclass{aastex631}

\shorttitle{Magnetars' X-ray versus spin-down luminosities}
\shortauthors{Ardavan}
\graphicspath{{./}{figures/}}
\usepackage{color}

\begin{document}

\title{Do magnetars' X-ray luminosities exceed their spin-down luminosities?}

\author{Houshang Ardavan}
\affiliation{Institute of Astronomy, University of Cambridge,\\
Madingley Road, Cambridge CB3 0HA, UK; {\url{ ardavan@ast.cam.ac.uk}}}



\begin{abstract}
The prevailing view that magnetars' X-ray luminosities exceed their spin-down luminosities is based on the assumption that the decay with distance of the flux of the X-rays received from magnetars obeys the inverse-square law.  The results presented here, of testing the hypothesis of independence of luminosities and distances of magnetars by means of the Efron--Petrosian statistic, do not uphold this assumption however: they imply that the observational data in the McGill Magnetar Catalog are consistent with the dependence $S\propto D^{-3/2}$ of the flux densities $S$ of these objects on their distances $D$ at substantially higher levels of significance than they are with the dependence $S\propto D^{-2}$.  These results are not incompatible with the requirements of the conservation of energy because the radiation process described in Ardavan (2021, MNRAS, 507, 4530), by which the superluminally moving current sheet in the magnetosphere of a neutron star is shown to generate the slowly decaying X-ray pulses, is intrinsically transient.  Once their over-estimation is rectified, the ratios of X-ray to spin-down luminosities of known magnetars turn out to be invariably lower than one.  A magnetar differs from other rotationally powered pulsars only in that it is observed along a privileged latitudinal direction relative to its spin axis: the closer is the line of sight to a direction in which the radiation from the current sheet is focused, the higher the frequency content and the lower the decay rate with distance of the observed radiation.  The outbursts characterizing the emission of a magnetar thus arise from sudden movements of its spin or magnetic axes. 
\end{abstract}


\section{Introduction}
\label{sec:introduction}

Magnetars are X-ray emitting neutron stats that are deemed too luminous to be rotationally powered: a notion that is based on the long spin periods and the relatively large spin-down rates of this set of neutron stars and the assumption that the decay of the observed emission with distance would necessarily obey the inverse-square law~\citep{Mereghetti2015,Turolla2015,Kaspi2017,Borghese2019,Esposito2021}.  This notion has led to the widely held view that such neutron stars are endowed with an ultra-strong magnetic field whose dissipation and instabilities power the emission received from them~\citep{Duncan1992} and underlie the flaring activity that characterizes this emission~\citep{Beloborodov2016}.  The purpose of the present paper is (i) to show that the observational data in the McGill Magnetar Catalog\footnote{\url{http://www.physics.mcgill.ca/~pulsar/magnetar/main.html}} do not uphold the notion on which the currently prevalent interpretation of these data is based and (ii) to present an alternative interpretation of the magnetar traits based on a recent study of the radiation by the superluminally moving current sheet in the magnetosphere of a neutron star~\citep{Ardavan2021,ArdavanHeuristic}: a study that has provided an all-encompassing explanation for the salient features of the radiation received from pulsars (its brightness temperature, polarization, spectrum and profile with microstructure and with a phase lag between the radio and gamma-ray peaks).

The radiation field generated by a constituent volume element of the current sheet in the magnetosphere of a neutron star embraces a synergy between the superluminal version of the field of synchrotron radiation and the vacuum version of the field of \v{C}erenkov radiation.  Once superposed to yield the emission from the entire volume of the source, the contributions from the volume elements of the current sheet that approach the observation point with the speed of light and zero acceleration at the retarded time interfere constructively and form caustics in certain latitudinal directions relative to the spin axis of the neutron star.  The waves that embody these caustics are more focused the further they are from their source: two nearby stationary points of their phases draw closer as their distance from their source increases and eventually coalesce at infinity.  As a result, flux densities of the pulses that are generated by this current sheet diminish with the distance $D$ from the star as $D^{-3/2}$ (rather than $D^{-2}$) along the latitudinal directions where they are most tightly focused~\citep[][Section 5.5]{Ardavan2021}.  By virtue of their extremely narrow peaks in the time domain, such pulses in addition have broad spectra that encompass X-ray and gamma-ray frequencies~\citep[][Table~1 and Section~5.4]{Ardavan2021}.  Analysis of the data in the second Fermi Catalog~\citep{Abdo2013} has already confirmed that the class of progenitors of this non-spherically decaying radiation includes the gamma-ray pulsars~\citep{Ardavan2022}.  As X-ray emitting neutron stars, magnetars are expected to be another member of this class. 

To see whether this expectation is supported by the observational data on magnetars~\citep{Olausen2014}, we analyse these data here on the basis of the fact that, in a statistical context, luminosities and distances of the sources of a given type of radiation represent two {\it independent} random variables.  A generalization of the nonparametric rank methods for testing the independence of two random variables~\citep{Hajek} to cases of truncated data, such as the flux-limited data on magnetars, is the method developed by~\citet{EF1992,EF1994}: a method that has been widely used in astrophysical contexts~\citep[see e.g.\ ][and the references therein]{Bryant}.   

From the raw data on fluxes $S$ and distances $D$ of magnetars in the McGill Online Magnetar Catalog and the candidate decay rates of flux density with distance ($S\propto D^{-\alpha}$ for various values of $\alpha$) we compile a collection of data sets on the prospective luminosities of magnetars (Section~\ref{sec:data}).  We then test the hypothesis of independence of luminosity and distance by evaluating the Efron--Petrosian statistic (Section~\ref{subsec:statistic}) for the data sets on prospective luminosities and magnetar distances with several different choices of the flux threshold (i.e.\ the truncation boundary below which the data set on $S$ may be regarded as incomplete).  The resulting values of the Efron--Petrosian statistic for differing values of $\alpha$ in each case determine the significance levels at which the hypothesis of independence of luminosity and distance (and hence a given value of $\alpha$) can be rejected (Section~\ref{subsec:results}).  

We also assess the effects of random and (if any) systematic errors in the estimates of distance and flux on the test results by means of a Monte Carlo simulation (Section~\ref{subsec:errors}).  Whether the size of the data set in the McGill Magnetar Catalog is sufficient for the purposes of the present rank analysis is ascertained in Section~\ref{subsec:size} where we augment this data set by the inclusion of its permissible permutations.  The alternative interpretation of the traits of magnetar emission, on the basis of the findings in~\citet{Ardavan2021,ArdavanHeuristic}, is presented in Section~\ref{sec:conclusion}.

\section{Observational data}
\label{sec:data}

\begin{figure*}
\centerline{\includegraphics[width=18cm]{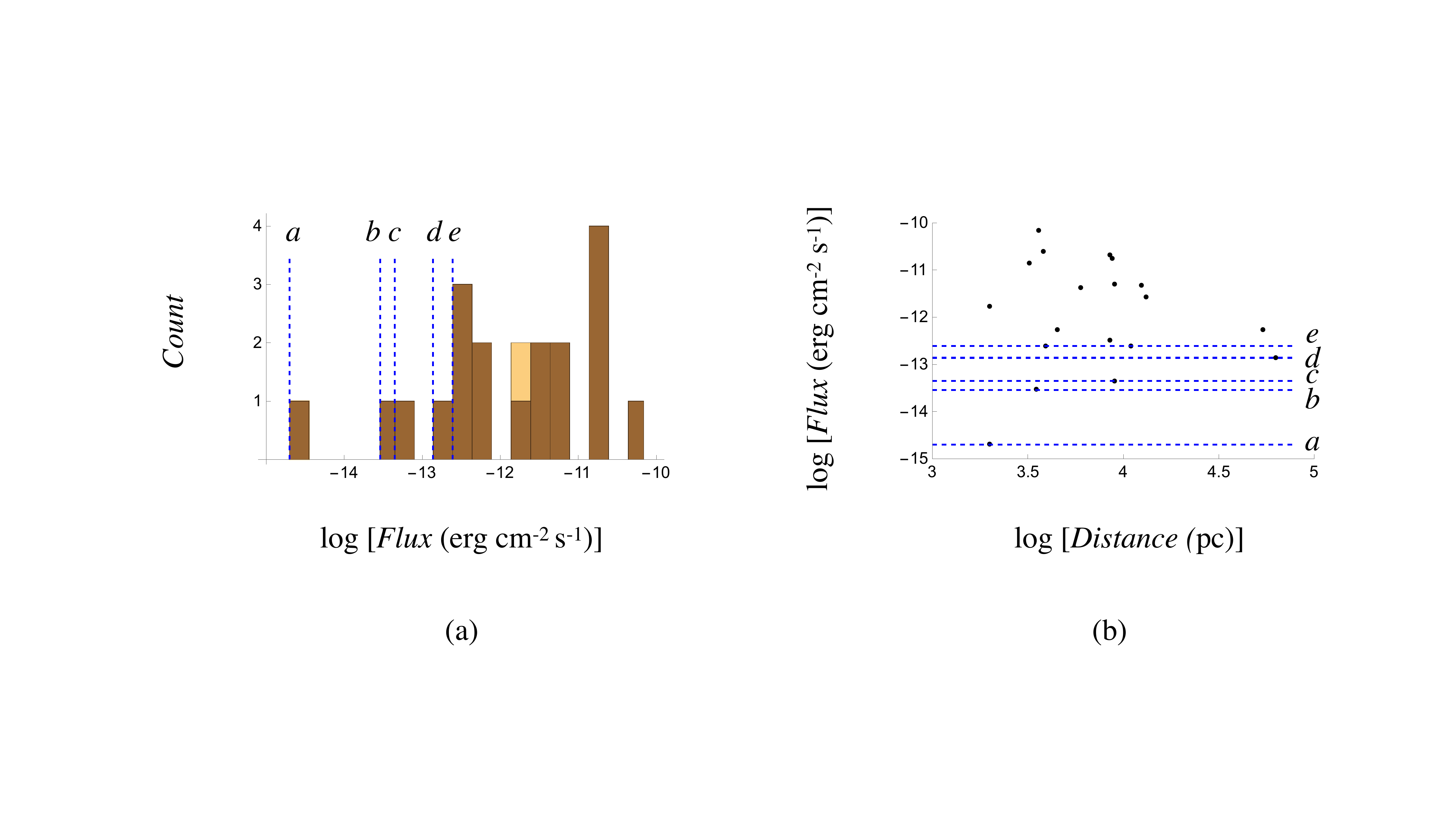}}
\caption{(a) Histogram of the $20$ magnetars whose fluxes are known.   The bins in darker brown contain the magnetars for which both fluxes and distances are known.  The broken lines $a$, $b$, $c$, $d$ and $e$ each designate a flux threshold below which the plotted data set may be incomplete.  (b) Distribution of logarithm of flux versus logarithm of distance for the $19$ magnetars in the darker brown bins.  The broken lines $a$, $b$, $c$, $d$ and $e$ designate the same flux thresholds as those shown in part (a). } 
\label{MF1}
\end{figure*}

The McGill Online Magnetar Catalog lists the observational data on $31$ magnetars of which $11$ have no flux estimates.  Histogram of the $20$ magnetars whose X-ray fluxes are known is shown in Fig.~\ref{MF1}a.  One of these has no distance estimate.  Logarithm of the X-ray flux (in units of erg cm$^{-2}$ s$^{-1}$) of each of the remaining $19$ magnetars is plotted versus the logarithm of its distance (in units of pc) in Fig.~\ref{MF1}b.  The dashed lines $a$, $b$, $c$, $d$ and $e$ in these figures each designate a flux threshold below which the plotted data set may be incomplete.

\begin{figure*}
\centerline{\includegraphics[width=15cm]{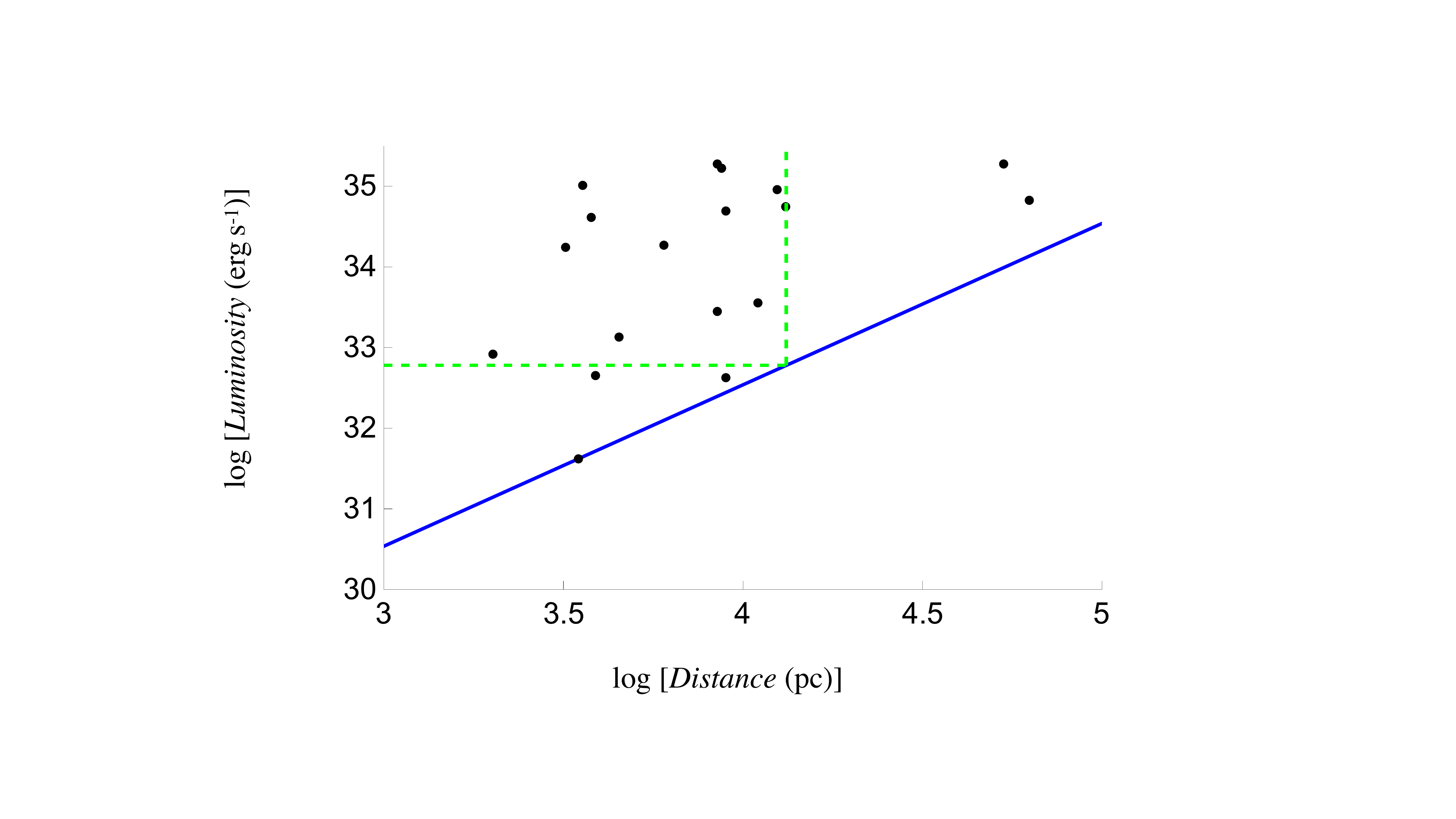}}
\caption{The luminosity-distance data set that follows from the flux-distance data set of Fig.~\ref{MF1}b and equation (\ref{E1}) for $\alpha=2$.  The solid line (in blue) is the image $L=L_{\rm th}(D)$ of the detection threshold that is marked by the dashed lines labelled $a$ in Figs~\ref{MF1}a and \ref{MF1}b.  Those elements of this data set that lie within (and on the boundary of) the rectangular area bounded by the vertical axis and the vertical and horizontal dashed lines (in green) comprise the set comparable to the element $(4.12, 34.75)$ on the vertical dashed line.} 
\label{MF2}
\end{figure*}

The isotropic X-ray luminosity of each magnetar is given, in terms of its flux density $S$ and its distance $D$, by 
\begin{equation}
L=4\pi \ell^2(D/\ell)^\alpha S,
\label{E1}
\end{equation}
where $\alpha=2$ if $S$ diminishes with distance as predicted by the inverse-square law and $\ell$ is a constant with the dimension of length whose value only affects the scale of $L$.  The corresponding distribution of the logarithm of $L$ (in units of erg s$^{-1}$) versus logarithm of $D$ (in units of pc) for the data set shown in Fig.~\ref{MF1}b is plotted in Fig.~\ref{MF2} for $\alpha=2$.  The solid line in Fig.~\ref{MF2} corresponds to the detection limit $a$ on the value of the flux density $S$ (see Fig.~\ref{MF1}).

\section{Testing the hypothesis that the data sets on luminosity and distance are independent}
\label{sec:test}
\subsection{The Efron--Petrosian rank statistic}
\label{subsec:statistic}

In this section we only outline the procedure by which the Efron--Petrosian statistic is calculated for two given data sets; lucid expositions of the theoretical basis of this procedure can be found in~\citet{EF1992}, \citet{Maloney1999} and~\citet{Petrosian2002}. 

If we let $S_{\rm th}$ stand for the threshold value of flux density, then the corresponding truncation boundary for the values of luminosity (e.g.\ that shown as a solid line in Fig.~\ref{MF2}) is given, according to equation~(\ref{E1}), by $L=L_{\rm th}(D)$ with
\begin{equation}
\log L_{\rm th}=\log[4\pi(3.085\times10^{18})^2\ell^{2-\alpha} S_{\rm th}]+\alpha\log D,
\label{E2}
\end{equation}
in which the numerical factor $3.085\times10^{18}$ converts the units of $D$ and $\ell$ from pc to cm.  The data set on luminosity is thus regarded as complete only in the sector $\log L\ge \log L_{\rm th}$ of the $(\log D,\log L)$ plane.

The set {\it comparable} to any given element $(\log D_i,\log L_i)$ of the bivariate distance-luminosity data set (such as the data set plotted in Fig.~\ref{MF2}) is defined to comprise all those elements for which
\begin{equation}
\log D\le \log D_i,\qquad i=1,\cdots n,
\label{E3}
\end{equation}  
and
\begin{equation}
\log L\ge\log[4\pi(3.085\times10^{18})^2\ell^{2-\alpha} S_{\rm th}]+\alpha\log D_i,
\label{E4}
\end{equation}
where $n$ is the number of elements in the part of the data set that is not excluded by the chosen flux threshold.  For instance, the set comparable to the data point $(4.12, 34.75)$ in Fig.~\ref{MF2} consists of the elements of the data set that lie within (and on the boundary of) the rectangular region delineated by the vertical axis and the broken lines coloured green in this figure.  We denote the number of elements in the set comparable to $(\log D_i,\log L_i)$ by $N_i$.

To determine the {\it rank} ($1\le R_i\le N_i$) of the element $(\log D_i,\log L_i)$, we now order the $N_i$ elements of its comparable set by the ascending values of their coordinates $\log L_i$ and equate $R_i$ to the position at which $(\log D_i,\log L_i)$ appears in the resulting ordered list.  Coordinates of the elements of the bivariate data set $(\log D,\log L)$ are in the present case all distinct.  So, a rank $R_i$ can be assigned to every element of this set unambiguously.  

The Efron--Petrosian {\it statistic} is given by
\begin{equation}
\tau=\frac{\sum_{i=1}^n(R_i-E_i)}{\sqrt{\sum_{i=1}^nV_i}},
\label{E5}
\end{equation}
in which
\begin{equation}
E_i=\textstyle{\frac{1}{2}}(N_i+1),\qquad V_i=\textstyle\frac{1}{12}(N_i^2-1).
\label{E6}
\end{equation}
Note that the value of $\tau$ is independent of that of the scale factor $\ell$ that appears in equation~(\ref{E1}): the value of the Efron--Petrosian statistic does not change if we make monotonically increasing transformations on the values of distance and/or flux~\citep{EF1992}.  

The hypothesis of independence is rejected when the value of $\tau$ falls in one of the tails of a Gaussian distribution whose mean is $0$ and whose variance is $1$.  More precisely, the hypothesis of independence is rejected when the value of
\begin{equation}
p=(2/\pi)^{1/2}\int_{\vert\tau\vert}^\infty\exp(-x^2/2){\rm d}x={\rm erfc}(\vert\tau\vert/\sqrt{2})
\label{E7}
\end{equation}
is smaller than an adopted significance level between $0$ and $1$, where ${\rm erfc}$ denotes the complementary error function.  If $\tau$ equals zero, for instance, then $p$ would assume the value one and the hypothesis in question cannot be rejected at any significance level.

\begin{figure*}
\centerline{\includegraphics[width=18cm]{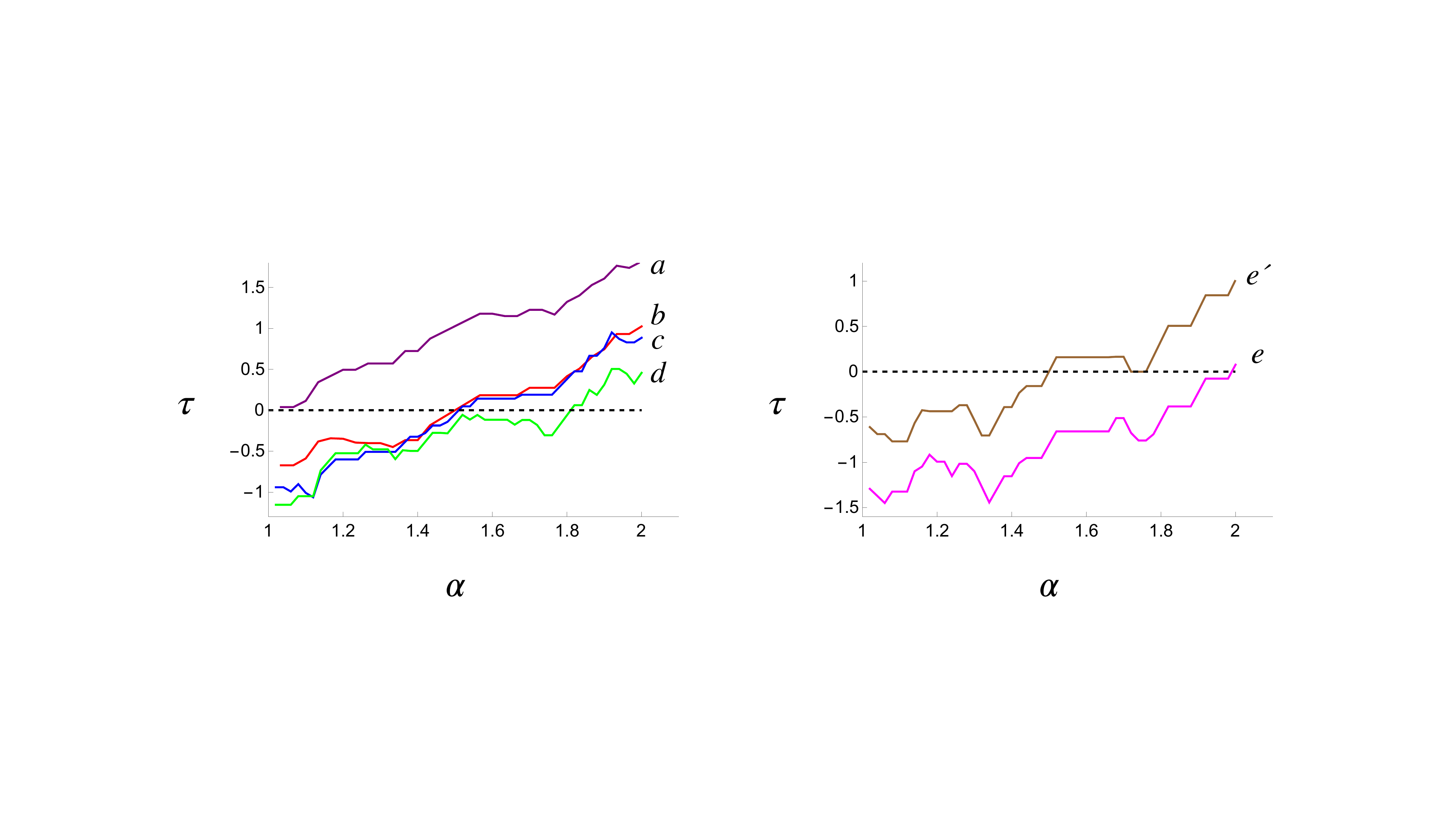}}
\caption{The Efron-Petrosian statistic $\tau$ as a function of the exponent $\alpha$ (in the dependence $S\propto D^{-\alpha}$ of flux density on distance) for the flux thresholds $a$, $b$, $c$, $d$ and $e$, designated by the dashed lines in Fig.~\ref{MF1}, and for the threshold $e^\prime$ close to $e$ (see Table~\ref{T1}).} 
\label{MF3}
\end{figure*}

\begin{figure*}
\centerline{\includegraphics[width=16cm]{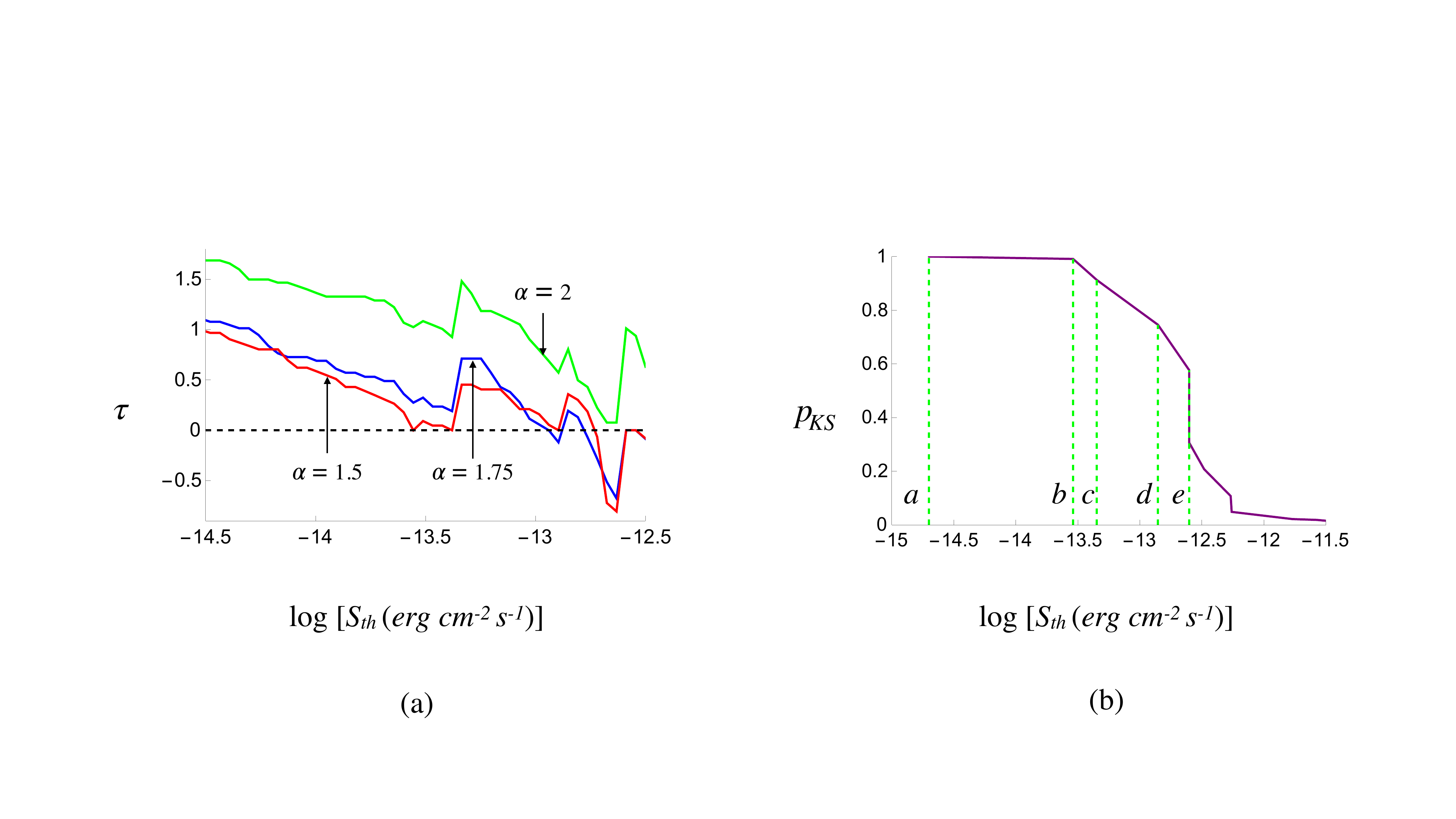}}
\caption{(a) The Efron--Petrosian statistic $\tau$ versus the logarithm of the flux threshold $S_{\rm th}$ for $\alpha=1.5$ (the red curve), $\alpha=1.75$ (the blue curve) and $\alpha=2$ (the green curve).  Note that $\tau$ vanishes when $\log S_{\rm th}=-13.56$, $-13.38$, $-12.89$, $-12.59$ or $-12.54$ for $\alpha=1.5$ and when $\log S_{\rm th}=-12.94$, $-12.87$, $-12.78$, $-12.59$ or $-12.54$ for $\alpha=1.75$.  (b)  The Kolmogorov--Smirnov statistic $p_{KS}$ for testing whether the $20$-element data set on fluxes shown in Fig.~\ref{MF1}a and the truncated versions of the 19-element data set shown in Fig.~\ref{MF1}b (from which the elements with fluxes $S < S_{\rm th}$ are eliminated) are drawn from the same distribution.  Note that while $p_{SK}$ has the values $1$, $0.991$, $0.913$ and $0.746$ for the thresholds $a$, $b$, $c$ and $d$, respectively, it jumps from $0.577$ to $0.306$ across threshold $e$.} 
\label{MF4}
\end{figure*}

\begin{table*}
\centering
\caption{The Efron--Petrosian statistic $\tau$ and its associated $p$ and $p_{\rm perm}$ values for $\alpha=2$ and $\alpha=1.5$ versus the flux threshold $S_{\rm th}$.  The listed values of $S_{\rm th}$ are designated in Figs.~\ref{MF1}, \ref{MF3} and \ref{MF4} by the letters $a$--$e^\prime$.}
\label{tab:landscape}
\begin{tabular}{lcccccc}
\hline
$\log [S_{\rm th}$ & $\tau\vert_{\alpha=2}$  & $p\vert_{\alpha=2}$ & $p_{\rm perm}\vert_{\alpha=2}$ & $\tau\vert_{\alpha=1.5}$ & $p\vert_{\alpha=1.5} $ & $p_{\rm perm}\vert_{\alpha=1.5} $\\
 (erg cm$^{-2}$ s$^{-1})]$ & & & & & &\\
\hline
 -14.70 ($a$)  & 1.818 & 0.069 & 0.092 & 1.027 & 0.304 & 0.304\\
 -13.54 ($b$) &  1.024 & 0.306 & 0.251 &  0     & 1 & 1\\
 -13.35 ($c$) & 0.885 & 0.376 & 0.331  & -0.047 & 0.963 & 0.943\\ 
 -12.86 ($d$) & 0.457 & 0.647 & 0.504 &-0.169 & 0.865 & 0.760\\
-12.61 ($e$) & 0.077 & 0.939 & 0.905 & -0.806 & 0.420 & 0.418\\
-12.59 ($e^\prime$) & 1.012 & 0.312 & 0.316 & 0 &1 & 1\\
\hline
\end{tabular}
\label{T1}
\end{table*}

\subsection{Test results}
\label{subsec:results}

We have evaluated the expression in equation~(\ref{E5}) for the Efron--Petrosian statistic $\tau$, as a function of the exponent $\alpha$ that appears in the expression for luminosity in equation (\ref{E1}), for the flux thresholds marked by the dashed lines in Figs.~\ref{MF1}a and \ref{MF1}b: for $\log S_{\rm th}=-14.70$ ($a$), $-13.54$ ($b$),  $-13.35$ ($c$), $-12.86$ ($d$) and $-12.61$ ($e$).  The number of elements of the present data set that are excluded by the thresholds $a$, $b$, $c$, $d$ and $e$ are $0$, $1$, $2$, $3$ and $4$, respectively.  The corresponding graphs of $\tau$ versus $\alpha$ for these thresholds are plotted in Fig.~\ref{MF3}.  To illustrate the sensitive dependence of $\tau$ on $S_{\rm th}$ in the vicinity of threshold $e$, we have also plotted $\tau$ versus $\alpha$ for the threshold $\log S_{\rm th}=-12.59$ neighbouring $e$ (labelled $e^\prime$) in Fig.~\ref{MF3}b.  The threshold labelled $e^\prime$ excludes $6$ elements of the present data set.

For threshold $a$, the resulting values of $\tau$ ($\tau=1.82$ when $\alpha=2$ and $\tau=1.03$ when $\alpha=3/2$) imply that the hypothesis of independence of luminosity and distance can be rejected at the significance levels $\ge 0.069$ when $\alpha=2$ and $\ge 0.304$ when $\alpha=3/2$.  The fact that neither values of $\tau$ are close to zero confirms that, as emphasized by~\citet{Bryant} in the context of gamma-ray bursts, the value of flux below which the data are incomplete lies closer to the peak of the histogram in Fig.~\ref{MF1}a than does the detection threshold $a$.  

For other thresholds, the resulting values of $\tau$ and their associated $p$-values are listed in Table~\ref{T1} in the cases where $\alpha=2$ or $\alpha=1.5$.  These values of $\tau$ and $p$ show that whichever of the thresholds $a$, $b$, $c$ or $d$, are chosen, there is always a wide range of significance levels at which the present hypothesis of independence can be rejected in the case of $\alpha=2$ but not in the case of $\alpha=1.5$.  We shall see that the exceptional outcome of the test in the case of threshold $e$ is radically modified once account is taken of the sharp changes in $\tau$ and in $p_{KS}$ (defined below) that result from slight variations in $\log S_{\rm th}$ for $e$ (see Figs.~\ref{MF3} and \ref{MF4}) and in the coordinates $(\log D_i,\log L_i)$ of data points within their error bars (see Fig.~\ref{MF6}).  

While the values of $\log S_{\rm th}$ for the thresholds $e$ and $e^\prime$ only differ by $0.02$, the difference between the corresponding values of $\tau\vert_{\alpha=1.5}$ for these two thresholds exceeds $0.8$ (see Table~\ref{T1}).  Moreover, the Kolmogorov--Smirnov test shows that the probability $p_{KS}$ that the truncated data set with the elements $S > S_{\rm th}$ and the uncut $20$-element data set depicted in Fig.~\ref{MF1}a are drawn from the same distribution sharply drops from $0.577$ to $0.306$ across $\log S_{\rm th}=-12.61$, i.e. across threshold $e$ (see Fig.~\ref{MF4}b).  The results of the Kolmogorov--Smirnov test depicted in Fig.~\ref{MF4}b also indicate that any truncated data set for which $\log S_{\rm th} > -12.61$ is significantly less likely to have the same origin as the uncut $20$-element data set than do those with the thresholds $b$, $c$, $d$ and $e$.  It is essential that the observationally obtained data set and the truncated part of it that lies above the chosen flux threshold could be regarded as drawn from the same distribution: from the unknown distribution that is complete over all values of the flux density.  This requirement sets a limit on how high the chosen value of the flux threshold can be. 

Given the rapid variations of $\tau$ across the flux thresholds $c$, $d$, $e$ and $e^\prime$ (see Fig.~\ref{MF4}a), the question that needs to be addressed next is: how resilient are the results listed in Table~\ref{T1} against the uncertainties in the estimates of magnetars' fluxes and distances?

\subsection{The effect of observational errors on the test results}
\label{subsec:errors}

Even if present, a purely systematic error in the estimates of distance and/or flux would not alter the results reported in Section~\ref{subsec:results}: the value of the Efron--Petrosian statistic does not change if we make monotonically increasing transformations on the values of distance and/or flux~\citep{EF1992}.  For instance, the dependence of $\tau$ on $\alpha$ (depicted by the curves in Fig.~\ref{MF3}) remains exactly the same if the distances in the data set shown in Fig.~\ref{MF1}b are all multiplied by a positive factor.  The McGill Magnetar Catalog lists the random errors in the estimates of distances of $14$ and of fluxes of $11$ of the $19$ data points plotted in Fig.~\ref{MF1}b.  In this section we perform a Monte Carlo simulation with $10^3$ random samplings in the case of each of the flux thresholds $b$, $c$, $d$, $e$ and $e^\prime$ to assess the effect of the listed random errors on the test results. 

\begin{table*}
\centering
\caption{Modified version of Table~\ref{T1} in which observational errors are taken into account.  The values listed in this table are based on the mean of the distribution of the Efron--Petrosian statistic $\tau$ for each threshold $S_{\rm th}$ (see Figs.~\ref{MF5} and \ref{MF6})} 
\label{T2}
\begin{tabular}{lcccccc}
\hline
$\log [S_{\rm th}$ & ${\bar\tau}\vert_{\alpha=2}$  & ${\bar p}\vert_{\alpha=2}$ & ${\bar p}_{\rm perm}\vert_{\alpha=2}$ & ${\bar\tau}\vert_{\alpha=1.5}$ & ${\bar p}\vert_{\alpha=1.5}$ & ${\bar p}_{\rm perm}\vert_{\alpha=1.5}$\\
 (erg cm$^{-2}$ s$^{-1})]$ & & & & & &\\
\hline
 -14.70 ($a$)  & 1.936 & 0.053 & 0.075 & 1.176 & 0.239 & 0.240\\
 -13.54 ($b$) &  1.259 & 0.208 & 0.205 &  0.296 & 0.767 & 0.683\\
 -13.35 ($c$) & 1.229 & 0.219 & 0.200  & 0.392 & 0.695 & 0.624\\ 
 -12.86 ($d$) & 0.440 & 0.660 & 0.513 &-0.016& 0.987 & 0.977\\
-12.61 ($e$) & 0.413 & 0.679 & 0.540 & -0.185 & 0.853 & 0.831\\
-12.59 ($e^\prime$) & 0.441 & 0.659 & 0.501 & -0.180 &0.857 & 0.883\\
\hline
\end{tabular}
\end{table*}

The values of distances and fluxes for which the observational errors are known are listed in the McGill Magnetar Catalog each with a positive ($\sigma_+$) and a negative ($\sigma_-$) uncertainty~\citep[see][]{Olausen2014}.  The distribution of the error in the value $\mu$ of each one of the listed variables can accordingly be modelled by an asymmetric Gaussian probability density:
\begin{equation}
f(x)= \sqrt{\frac{2}{\pi}}\left[\sigma_++{\rm erf}\left(\frac{\mu}{\sqrt{2}\sigma_-}\right)\sigma_-\right]^{-1}\left\{\exp\left[-\frac{(x-\mu)^2}{2\sigma_-^2}\right]{\rm H}(\mu-x)+\exp\left[-\frac{(x-\mu)^2}{2\sigma_+^2}\right]{\rm H}(x-\mu)\right\},\quad 0<x<\infty,
\label{E8}
\end{equation}
where erf and H denote the error function and the Heaviside step function, respectively.  Equation~(\ref{E8}) describes half a Guassian distribution with the mean $\mu$ and the standard deviation $\sigma_-$ in $0 \le x \le \mu$ and half a Gaussian distribution with the same mean but the standard deviation $\sigma_+$ in $\mu \le x <\infty$.  (For some of the listed variables, $\sigma_+$ and $\sigma_-$ have the same value.)

The sampling domain of the Monte Carlo method we use consists of the collection of the intervals $\mu-\sigma_- \le x \le\mu+ \sigma_+$ which enclose   
each of the two coordinates (whose listed value we have denoted by $\mu$) of the elements of the data set shown in Fig.~\ref{MF1}b.  In a given sampling, the values of distance and flux for each element of the data set in Fig.~\ref{MF1}b are replaced by two randomly chosen values of $x$ from the probability distribution $f(x)$ over the interval $\mu-\sigma_- \le x \le \mu+\sigma_+$, where $\mu$ and $\sigma_\pm$ are the parameters of the relevant coordinate (either distance or flux) of the data point in question as they appear in the McGill Magnetar Catalog.  The modified data set thus obtained in a given sampling is then used, in conjunction with a choice of the flux threshold $S_{\rm th}$, to calculate the Efron--Petrosian statistic $\tau$ for a given value of $\alpha$.  

\begin{figure*}
\centerline{\includegraphics[width=16cm]{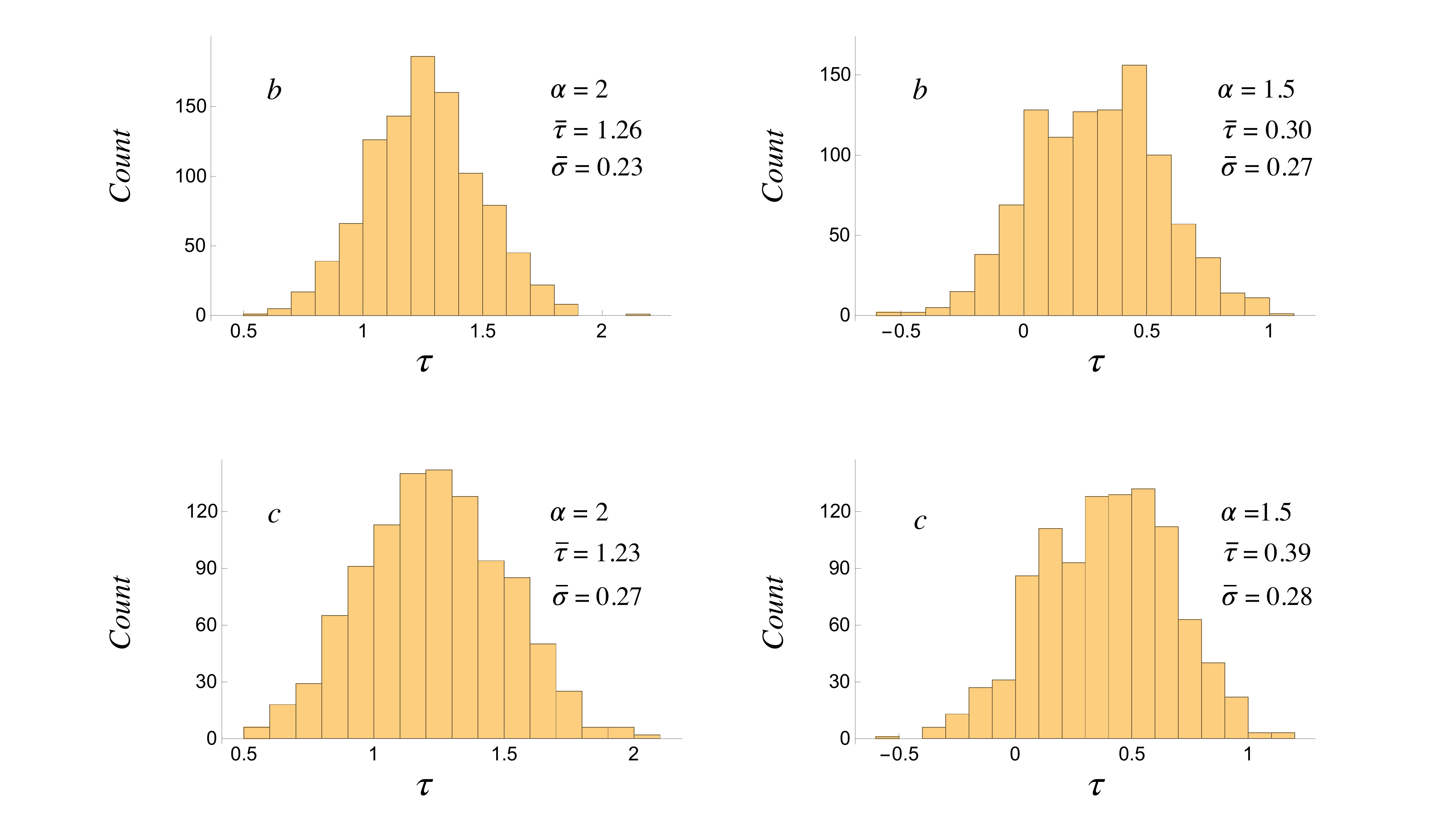}}
\caption{Histograms of the distributions of the Efron--Petrosian statistic $\tau$ for $\alpha=2$ and $\alpha=1.5$ in the cases of the flux thresholds $b$ (the upper figures) and $c$ (the lower figures).  The mean and the standard deviation of each distribution are denoted by ${\bar\tau}$ and ${\bar\sigma}$, respectively.} 
\label{MF5}
\end{figure*}

\begin{figure*}
\centerline{\includegraphics[width=17cm]{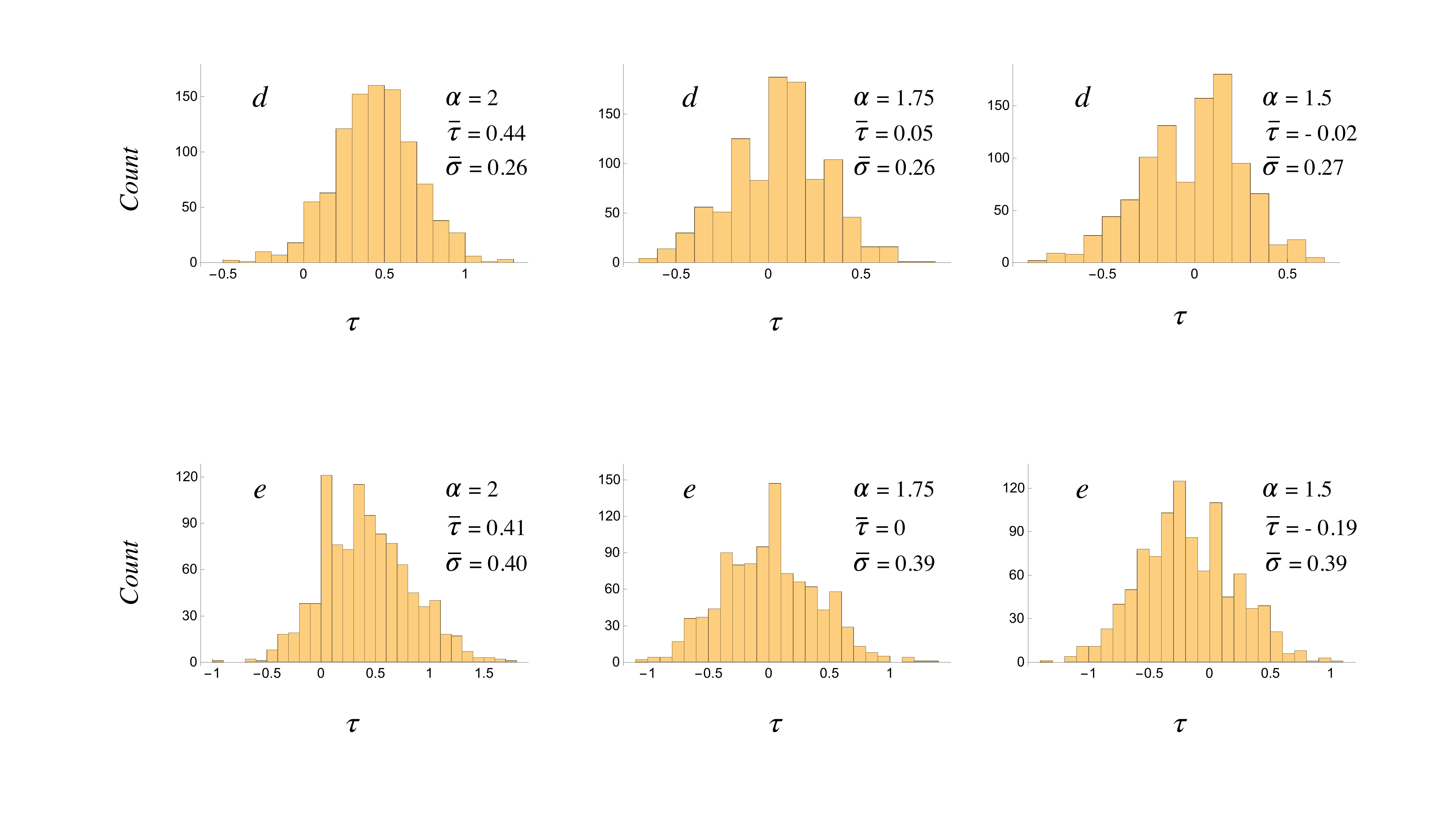}}
\caption{Histograms of the distributions of the Efron--Petrosian statistic $\tau$ for $\alpha=2$, $\alpha=1.75$ and $\alpha=1.5$ in the cases of the flux thresholds $d$ (the upper figures) and $e$ (the lower figures).  The mean and the standard deviation of each distribution are denoted by ${\bar\tau}$ and ${\bar\sigma}$, respectively.} 
\label{MF6}
\end{figure*}

Repeating this procedure $10^3$ times for each threshold and aggregating the resulting sets of values of $\tau$ for $\alpha=2$, $\alpha=1.75$ and $\alpha=1.5$, we arrive at distributions of the Efron--Petrosian statistic whose scatters describe the effect of random errors (see Figs.~\ref{MF5} and \ref{MF6}).  The mean values of these distributions (denoted ${\bar\tau}$) and their associated $p$-values (denoted ${\bar p}$) listed in Table~\ref{T2} show that -- with the exception of those for threshold $e$ -- the uncertainties in the estimates of magnetars' distances and fluxes, though bringing about a spreading of the values of $\tau$, point to the same conclusions as those reached on the basis of the test results depicted in Fig.~\ref{MF3} and listed in Table~\ref{T1}.  For thresholds $a$, $b$, $c$ and $d$, the ratios ${\bar p}\vert_{\alpha=1.5}/{\bar p}\vert_{\alpha=2}$ of the values of ${\bar p}$ for $\alpha=1.5$ and for $\alpha=2$ do not significantly differ from those of the corresponding ratios of the values of $p$ (cf. Tables~\ref{T1} and \ref{T2}).  

For thresholds $e$ and $e^\prime$, the values of ${\bar p}\vert_{\alpha=1.5}/{\bar p}\vert_{\alpha=2}$ replacing the discordant values of $p\vert_{\alpha=1.5}/p\vert_{\alpha=2}$ across $\log S_{\rm th}=-12.60$ are comparable, so that the discontinuity in the value of $\tau$ (encountered in Fig.~\ref{MF3} and in Table~\ref{T1}) is removed by the inclusion of the observational uncertainties: unlike the curve in Fig.~\ref{MF3} for threshold $e^\prime$ which differs from that for threshold $e$ by a large margin, distributions of the scattered values of $\tau$ for $e^\prime$ turn out to be practically the same as those that are depicted in Fig.~\ref{MF6} for $e$.  The exceptional value $10.47$ of the ratio $\vert\tau\vert_{\alpha=1.5}/\vert\tau\vert_{\alpha=2}$ implied by the curve for threshold $e$ in Fig.~\ref{MF3}  is thus reduced to a value $\vert{\bar\tau}\vert_{\alpha=1.5}/\vert{\bar\tau}\vert_{\alpha=2}=0.448$ comparable to those found in the case of other thresholds.  Moreover, as suggested by the proximity of the curves for $\alpha=1.75$ and $\alpha=1.5$ in Fig.~\ref{MF4}a, the values of ${\bar\tau}\vert_{\alpha=1.75}$ and ${\bar\tau}\vert_{\alpha=1.5}$ (and hence those of ${\bar p}\vert_{\alpha=1.75}$ and ${\bar p}\vert_{\alpha=1.5}$) are hardly distinguishable in the cases of thresholds $d$, $e$ and $e^\prime$ (see Fig.~\ref{MF6}). 

For every one of the thresholds $a$, $b$, $c$, $d$, $e$ and $e^\prime$, therefore, there is a wide range of significance levels at which the hypothesis of independence of luminosity and distance can be rejected in the case of $\alpha=2$ but not in the case of $\alpha=1.5$ (or in the case of $1.5\le\alpha\le1.75$ for thresholds $d$, $e$ and $e^\prime$).

\subsection{The effect of the limited size of the data set on the test results}
\label{subsec:size}

The only step in the analysis described in Section~\ref{subsec:statistic} at which the size of the data set is assumed to be large is where the resulting value of the Efron--Petrosian statistic $\tau$ is appraised by comparison with a normalized Gaussian distribution, i.e.\ the very last step at which the $p$ value associated with $\tau$ is calculated by means of equation~(\ref{E7}).  By relaxing that assumption, one can in fact apply the Efron--Petrosian method to data sets of any size: the normalized Gaussian distribution should be replaced, in the general case, by the normalized permutation distribution of the Efron--Petrosian statistic for the data set~\citep[see][Section 2.2]{EF1992}.

A permissible permutation of a truncated luminosity-distance data set, such as that shown in Fig.~\ref{MF2}, is obtained by replacing an element $(\log D_i, \log L_i)$ of that data set by any one of the $N_i$ elements that constitute the set comparable to $(\log D_i, \log L_i)$.  The number of permuted data sets thus obtained from an original $n$-element data set is therefore given by $N=\sum_1^n N_i$.  The value of $N$ for the $17$-element data set corresponding to threshold $c$, for example, is $134$ when $\alpha=2$ and $139$ when $\alpha=3/2$.

If we calculate the Efron--Petrosian statistic $\tau$ for each one of the $N$ permuted data sets (corresponding to a given threshold and a given value of $\alpha$) and normalize the distribution of the resulting values of $\tau$ (by replacing each $\tau$ with $(\tau - {\bar\tau})/\sigma$, where ${\bar\tau}$ and $\sigma$ are the mean and standard deviation of the distribution), we arrive at a normalized probability distribution function representing the permutation distribution of $\tau$.  In the case of a data set with a limited number of elements, the observed value $\tau_{\rm obs}$ of $\tau$ (i.e.\ the value of $\tau$ determined by the original data set) should be appraised by comparison with this permutation distribution of $\tau$ instead of the normalized Gaussian distribution used at the end of Section~\ref{subsec:statistic}.  The size of the area that lies under a normalized permutation distribution in $-\infty<\tau\le-\vert\tau_{\rm obs}\vert$ and $\vert\tau_{\rm obs}\vert\le\tau<\infty$, here denoted by $p_{\rm perm}$, plays the role of the $p$-value defined in equation (\ref{E7})~\citep[see][Section 2.2]{EF1992}.

\begin{figure*}
\centerline{\includegraphics[width=16cm]{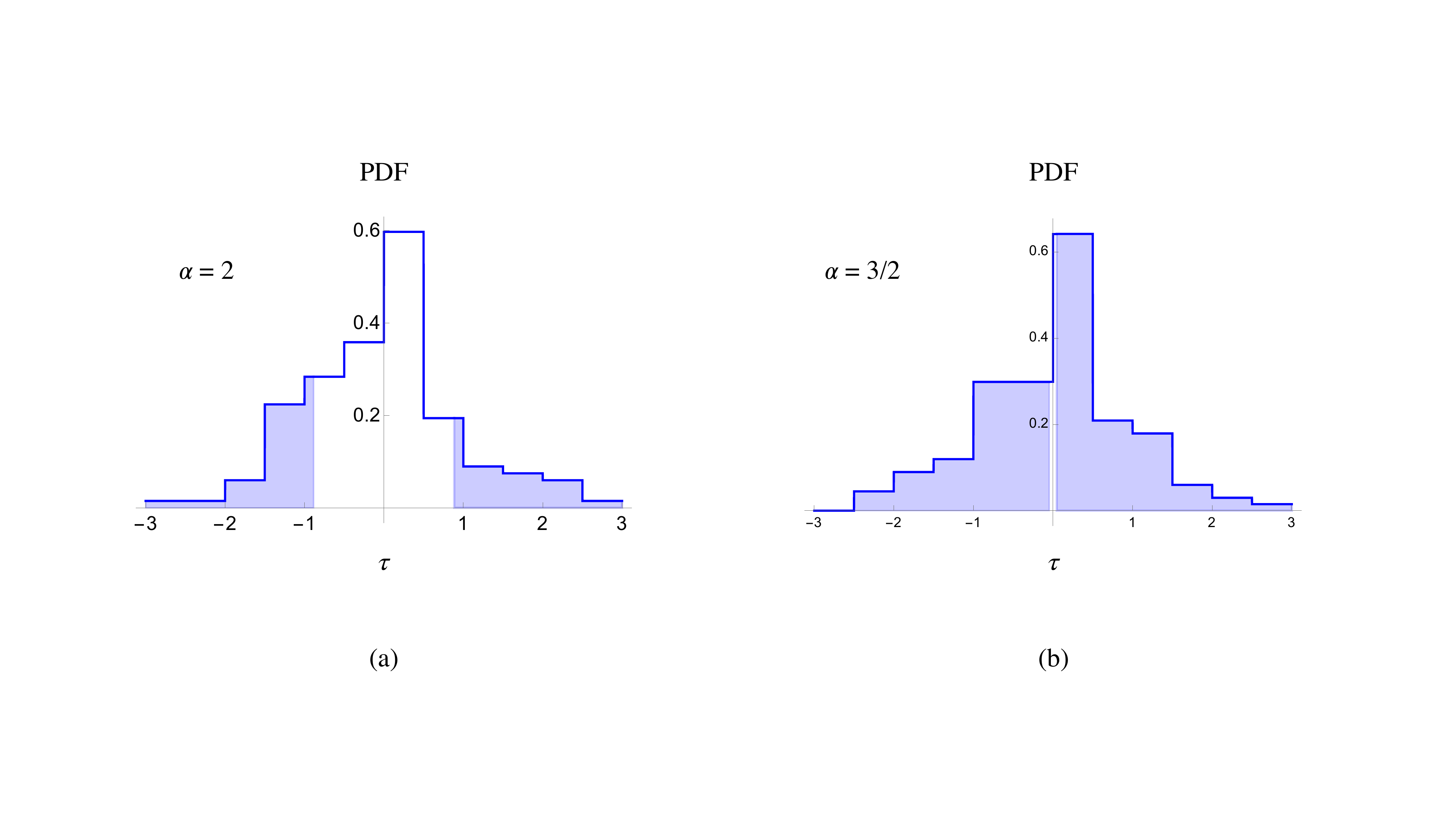}}
\caption{The normalized probability distribution functions of the Efron--Petrosian statistic for the observationally-determined $17$-element data set that corresponds to the flux threshold $c$.  (a) The permutation distribution of $\tau$ for $\alpha=2$ based on an augmented data set consisting of $134$ elements.  The shaded area under this curve yields $p_{\rm perm}=0.331$ for $\tau_{\rm obs}=0.885$, i.e.\ for the value of $\tau$ that is selected by $\alpha=2$ and the blue curve designating threshold $c$ in Fig.~\ref{MF3}.   (b) The permutation distribution of $\tau$ for $\alpha=3/2$ based on an augmented data set consisting of $139$ elements.  The shaded area under this curve yields $p_{\rm perm}=0.948$ for $\tau_{\rm obs}=-0.047$, i.e.\ for the value of $\tau$ that is selected by $\alpha=3/2$ and the blue curve designating threshold $c$ in Fig.~\ref{MF3}.}
\label{MF7}
\end{figure*}

The normalized probability distribution functions of the Efron--Petrosian statistic for the flux threshold $c$ and the values $2$ and $3/2$ of $\alpha$ are shown in Fig.~\ref{MF7}.  The mean and standard deviation of each of the shown distributions are given by $0$ and $1$, respectively.  The parameters $({\bar\tau},\sigma)$, i.e.\ the mean and standard deviation, of the un-normalized versions of these distributions have the values $(0.831, 0.328)$ for $\alpha=2$ and $(-0.052, 0.338)$ for $\alpha=3/2$.  The shaded regions in this figure, whose areas determine $p_{\rm perm}\vert_{\alpha=2}$ and $p_{\rm perm}\vert_{\alpha=3/2}$, are here sketched over the intervals $-3<\tau\le-\vert\tau_{\rm obs}\vert$ and $\vert\tau_{\rm obs}\vert\le\tau<3$, where $\tau_{\rm obs}=0.885$ for $\alpha=2$ and $\tau_{\rm obs}=-0.047$ for $\alpha=3/2$ (see Table~\ref{T1}). 

Figure~\ref{MF7}, and its counterparts for other flux thresholds, show that taking account of the limited size of the present data set does not appreciably alter the outcomes of the test results obtained before: $p\vert_{\alpha=2}=0.376$ and $p\vert_{\alpha=3/2}=0.963$ are respectively replaced by $p_{\rm perm}\vert_{\alpha=2}=0.331$ and $p_{\rm perm}\vert_{\alpha=3/2}=0.948$ (see Table~\ref{T1}), and ${\bar p}\vert_{\alpha=2}=0.219$ and ${\bar p}\vert_{\alpha=3/2}=0.695$ are respectively replaced by $p_{\rm perm}\vert_{\alpha=2}=0.200$ and $p_{\rm perm}\vert_{\alpha=3/2}=0.624$ (see Table~\ref{T2}).

\section{Discussion}
\label{sec:conclusion}

The conclusion to be drawn from the above results is that the observational data in the McGill Magnetar Catalog are consistent with the dependence $S\propto D^{-3/2}$ of the flux densities $S$ of magnetars on their distances $D$ at substantially higher levels of significance than they are with the dependence $S\propto D^{-2}$.  The violation of the inverse-square law encountered here is not incompatible with the requirements of the conservation of energy because the radiation process by which the superluminally moving current sheet in the magnetosphere of a neutron star generates the slowly decaying X-ray pulses is intrinsically transient.  Temporal rate of change of the energy density of the radiation generated by this process has a time-averaged value that is negative (instead of being zero as in a steady state) at points where the envelopes of the wave fronts emanating from the constituent volume elements of the current sheet are cusped~\citep[][Fig.~1 and Section~3.2]{Ardavan2021}.  The difference in the fluxes of power across any two spheres centered on the star is thus balanced by the change with time of the energy contained inside the shell bounded by those spheres~\citep[][ Appendix~C, where this is demonstrated for each high-frequency Fourier component of a superluminally rotating source distribution]{Ardavan_JPP}.

X-ray luminosities of magnetars are over-estimated when the decay of their flux density $S$ is assumed to obey the inverse-square law $S\propto D^{-2}$ instead of $S\propto D^{-3/2}$ by the factor $(D/\ell)^{1/2}$ (see equation~\ref{E1}).  The value of the scale factor $\ell$ is of  the same order of magnitude as the values of the light-cylinder radii of the central neutron stars of magnetars~\citep[][Section~5.5]{Ardavan2021}.  Hence, the factor by which the luminosity of a $5$~s magnetar at a distance of $8$~kpc is thus over-estimated is approximately $10^6$.  Once this is multiplied by the ratio $\sim10^{-2}$ of latitudinal beam-widths of magnetars and radio pulsars (implied by the fraction of known neutron stars that are identified as magnetars), we obtain a value of the order of $10^4$ for the over-estimation factor: a result that implies that the values of the correctly-estimated luminosities of magnetars are invariably lower than those of the spin-down luminosities of these objects (cf.\ the McGill Magnetar Catalog).

Contrary to the prevailing view~\citep{Mereghetti2015,Turolla2015,Kaspi2017,Borghese2019,Esposito2021}, therefore, the energetic requirements of magnetars are not different from those of rotationally-powered pulsars.  A magnetar differs from a rotationally-powered pulsar only in that it is observed along a privileged latitudinal direction relative to its spin axis.  The closer is the line of sight to a direction in which the radiation from the current sheet in the magnetosphere of a neutron star is focused, the higher the frequency content and the lower the decay rate with distance of the observed radiation~\citep[][Section~5.5]{Ardavan2021}.  If the frequency at which the spectrum of the radiation from a given magnetar peaks is lower than that at which the spectra of most gamma-ray pulsars do, it is because the line of sight to that magnetar does not lie as close to one of the privileged directions in question as the lines of sight to gamma-ray pulsars do.   This is borne out by the fact that (unlike the data in the second Fermi Catalog on gamma-ray pulsars which are consistent at high significance levels only with $\alpha=1.5$) the data in the McGill Magnetar Catalog are consistent with both $\alpha=1.5$ and $\alpha=1.75$ at comparable levels of significance for certain thresholds (see Fig.~\ref{MF6}).

The above rates of decay apply to the flux densities of both the hard and the soft components of the X-ray emission from a magnetar.  The spectrum of the radiation that is generated by the magnetospheric current sheet is described by an oscillatory function the amplitude of whose oscillations has an algebraic dependence on frequency with a wide range of power-law indices~\citep[][Table~2 and Fig.~18]{Ardavan2021}.  When observed over a limited frequency band in which only a single oscillation of its distribution is detectable, this spectrum resembles a distorted version of the black body spectrum.  However, both the oscillations of this spectrum and the power-law decay of their amplitudes are generated by the same non-thermal radiation process and are both described by the same Airy functions: functions that are emblematic of caustics~\citep[][Section 5.3]{Ardavan2021}. The observational data over a wider range of frequencies, collected from different space telescopes, indeed confirm the oscillatory nature of the spectrum of magnetars~\citep[see][]{denHartog2008}. 

For any given value of the angle between the magnetic and spin axes of the neutron star, there are four critical colatitudes~\citep[denoted by $\theta_{P1S}$, $\theta_{P2S}$, $\pi-\theta_{P1S}$ and $\pi-\theta_{P2S}$ in][]{Ardavan2021} with respect to the spin axis of the star along which the flux density of the radiation decays non-spherically.   The gradual change in the rate of decay of flux density with distance, from $D^{-3/2}$ to $D^{-2}$, away from a critical colatitude takes place over a latitudinal interval of the order of a radian.  But the latitudinal width of the tightly focused part of this radiation beam whose flux density decreases as $D^{-3/2}$ with distance $D$ has a much lower value: it is of the order of $(D/R_{lc})^{-1}$, where $R_{lc}$ is the radius of the star's light cylinder~\citep[][Section 5.5]{Ardavan2021}.  Hence, the solid angle centered on the neutron star within which the flux density of the present radiation decays at a significantly lower rate than that predicted by the inverse-square law is only a small fraction of $4\pi$.  

It follows that our lines of sight to only a small fraction of the total number of known neutron stars are expected to coincide with the privileged directions along which the flux density of the radiation we observe would decay as $D^{-3/2}$.  This is borne out by the fact that the known X-ray and gamma-ray emitting neutron stars comprise a small fraction of all observed pulsars.  

The closer is the line of sight to a privileged direction relative to the spin axis of the neutron star, the higher the amplitude of the observed pulse and the broader its frequency spectrum.  Moreover, the pulse profiles depicted in Figs~8--16 of~\citet{Ardavan2021} show that mode changes arise, in the case of the current-sheet emission, from changes in the angle between the line of sight and the magnetic or rotation axes of the central neutron star.  The fact that magnetar outbursts are often accompanied by timing anomalies such as glitches or mode changes~\citep{Camilo2007,Borghese2019,Champion2020,Rajwade2022} implies, therefore, that the flare activity in these objects is caused by movements of their magnetic or spin axes.  Gamma-ray or X-ray outbursts occur when, as a result of a sudden movement of the magnetic or the spin axis of the neutron star, one of the privileged directions along which the radiation is focused either swings past or oscillates across the line of sight.  The large factor by which the amplitude of the observed radiation rises during an outburst simply reflects the factor $(D/\ell)^{1/2}$ by which the decay rates of the flux density in directions close to and far from a critical latitude differ from one another (see equation~\ref{E1}).  

Sudden movements of the spin or magnetic axes and the gradual relaxation of these axes to an original or a different direction account not only for the sudden rises in flux density that characterize the observed outbursts but also for any differences between the post-outburst and the original quiescent states of a magnetar~\citep{Coti2020}.  The spectrum of the observed radiation is also expected to change during an outburst since the full widths at half maxima of the pulses that are generated by the current sheet in the time-domain sensitively depend on the angle between the line of sight and a direction in which these pulses are focused~\citep[][Sections 5.3 and 5.5]{Ardavan2021}.

Not all glitching neutron stars are observed to partake in flaring activity because not all glitches would result in the alignment of the line of sight with a direction in which the radiation is focused.  Such an alignment is more likely to occur when the line of sight already lies close to a privileged direction associated with the quiescent state of the neutron star, as in the cases of gamma-ray pulsars and magnetars whose flux densities decay more slowly than predicted by the inverse-square law also when they are in a quiescent state.

That the emission mechanism of magnetars is no different from that of pulsars is further evidenced by the discoveries of magnetars that emit radio waves and pulsars that undergo magnetar-like outbursts~\citep{Borghese2020,Israel2021,Rajwade2022}.  An example of a feature that would be expected to be common to most members of these two sets of neutron stars, if their emissions arise from their magnetospheric current sheets in both cases, is the phase lag between the low-frequency and high-frequency peaks of their pulse profiles~\citep[][Section~5.4]{Ardavan2021}.  Not only is this phase lag observed in the case of the majority of radio-loud gamma-ray pulsars~\citep{Abdo2013} but it is also displayed by the radio and X-ray outbursts of objects that are identified as magnetars~\citep{Gotthelf2019}.

A final remark is in order: the high strength normally attributed to the surface magnetic field of a magnetar is estimated by means of the formula for the Poynting flux of an obliquely rotating magnetic dipole.  As this formula has no relevance to the process by which the current sheet in the magnetosphere of a neutron star radiates~\citep{Ardavan2021}, the central neutron stars of magnetars need not be more strongly magnetized than those of normal pulsars.


\bibliography{Magnetars.bib}{}
\bibliographystyle{aasjournal}



\end{document}